\documentclass[fleqn,twoside,11pt]{article}

\usepackage{amssymb}
\usepackage{graphicx}
\usepackage{epsbox}
\usepackage{amsmath}

\begin{document}
\noindent



\vspace{3mm}

\begin{center}

{\Large\bf  Seesaw Mass Matrix Model of Quarks and Leptons }\\[.1in]
{\Large\bf with Flavor-Triplet Higgs Scalars}

\vspace{3mm}
{\bf Yoshio Koide}

{\it Department of Physics, University of Shizuoka, 
52-1 Yada, Shizuoka 422-8526, Japan\\
E-mail address: koide@u-shizuoka-ken.ac.jp}

\date{\today}
\end{center}

\begin{abstract}
In a seesaw mass matrix model $M_f = m_L M_F^{-1} m_R^\dagger$ 
with a universal structure of $m_L \propto m_R$, as the origin
of $m_L$ ($m_R$) for quarks and leptons, 
flavor-triplet Higgs scalars whose vacuum expectation values $v_i$
are proportional to the square roots of the charged lepton masses
$m_{ei}$, i.e. $v_i \propto \sqrt{m_{ei}}$,  are assumed.
Then, it is investigated whether such a model can explain the 
observed neutrino masses and mixings (and also quark masses and
mixings) or not.
\end{abstract}

\vspace{3mm}

{\large\bf 1 \ Introduction}

It is widely accepted that quarks and leptons 
are fundamental entities of the matter.
If it is true, the masses and mixings of the quarks and 
leptons will obey a simple law of nature, and we will be 
able to find a beautiful relation among those values. 
If we can find such a relation, 
it will make a breakthrough in the unified understanding 
of the quarks and leptons. 
%
As one of such phenomenological mass relations, the following
charged lepton mass formula \cite{Koide82,Koide83,Koide90}
$$
m_e+m_{\mu}+m_{\tau}=
\frac{2}{3}\left( \sqrt{m_e}+\sqrt{m_\nu}+\sqrt{m_{\tau}} 
\right)^2 , 
\eqno(1.1)
$$
has been known.
The formula (1.1) predicts the tau lepton mass value 
$$
m_{\tau}=1776.97 \ {\rm MeV},
\eqno(1.2)
$$
from the observed electron and muon mass values 
\cite{PDG04}, $m_e=0.51099892$ MeV and 
$m_{\mu}=105.658369$ MeV. 
The predicted value (1.2) is in excellent agreement 
with the observed value \cite{PDG04} 
$m_{\tau}=1776.99^{+0.29}_{-0.26}$ MeV.  
This excellent agreement seems to be beyond a matter
of accidental coincidence,
so that we should consider the origin of the mass 
formula (1.1) seriously.
Up to the present, the theoretical basis of 
the mass formula (1.1) is still not clear. 
However, although it is still important to pursue the origin of the
relation (1.1), in the present paper, another approach will be
taken: We assume the so-called universal seesaw mass matrix model
\cite{UnivSeesaw} for an explanation of the
charged lepton mass relation (1.1) and it is investigate
whether the seesaw mass matrix model can also explain 
the observed quark and neutrino masses and mixings or not 
when the mass matrix parameters are 
settled by the observed charged lepton masses.

In order to obtain a clue to a unified description of the
quark and lepton mass matrices, let us see the 
phenomenological features of the relation (1.1).
The charged lepton mass formula (1.1) has 
the following peculiar features:

\noindent(a) The mass formula is described in terms of the 
root squared masse $\sqrt{m_{ei}}$.

\noindent(b) The mass formula is invariant under the exchanges 
$\sqrt{m_{ei}} \leftrightarrow \sqrt{m_{ej}}$.

\noindent(c) The formula gives a relation between mass ratios 
$\sqrt{m_e / m_{\mu}}$ and $\sqrt{m_{\mu} / m_{\tau}}$, 
whose 
behaviors under the renormalization group equation (RGE) 
effects are different from each other. 
Therefore, the formula (1.1) is not invariant 
under the RGE effects. 
The formula is well satisfied at a low energy scale 
rather than at a high energy scale. 

If we take the feature (c) seriously, we must abandon
the idea that the mass spectrum originates in the structure
of the Yukawa coupling constants $Y_e$, because, in general,
the Yukawa coupling constants are influenced by the
renormalization group equation (RGE) effects.
Even if the mass spectrum satisfies the relation (1.1)
at a unification energy scale $\mu= M_X$, the mass
spectrum at a low energy scale will satisfy the relation
(1.1) no longer.
We should consider that the Yukawa coupling constant $Y_e$ 
has a unit matrix form which is not affected by RGE effects.
Instead, we consider that the mass spectrum originates the
vacuum expectation values (VEVs) $v_i$ of three Higgs
scalars $\phi_i$ ($i=1,2,3$) at a low energy scale.

The feature (a) suggests that the charged lepton 
mass spectrum does not originate in the Yukawa coupling 
structure at the tree level, 
but it is given by a bilinear form 
on the basis of some mass-generation mechanism. 
For example, 
in Refs.~\cite{Koide90, KF96, KT96, Koide99}, 
a seesaw-like mechanism \cite{UnivSeesaw} has been assumed: 
$M_e=mM^{-1}_E m^{\dagger}$,
where $M_E$ is a mass matrix of hypothetical heavy leptons.
As suggested from the feature (c), we consider that the matrix
$m$ is given by $m_{ij} = \delta_{ij} v_j$.

The feature (b) suggests that the theory is 
invariant under a permutation symmetry S$_3$ \cite{S3}.
We will adopt an idea that what is essential 
is not a structure of the Yukawa coupling constants, 
but a structure of the vacuum expectation values (VEVs) 
of flavor-triplet (3-family) Higgs scalars 
\cite{Koide90,KT96,Koide99}. 
In this idea, the VEVs $v_i$ satisfies the relation
$$
v_1^2 + v_2^2 + v_3^2 = \frac{2}{3}
 \left( v_1 + v_2 + v_3 \right) ^2.
\eqno(1.3)
$$
(For the derivation of the relation (1.3), for example, see
Ref.\cite{Koide06-S3}.)
Then, the charged lepton mass relation (1.1) is understood from
a seesaw-like mechanism
$$
M_e = m_L M_E^{-1} m_R^\dagger ,
\eqno(1.3)
$$
$$
m_L = \frac{1}{\kappa} m_R = y_e {\rm diag}(v_1, v_2, v_3) ,
\eqno(1.4)
$$
$$
M_E = \mu_E {\bf 1} \equiv \mu_E {\rm diag}(1,1,1),
\eqno(1.5)
$$
where $m_L$ and $m_R$ are Dirac mass matrices for fermions
$(\bar{e}_L, E_E)$ and $(\bar{E}_L, e_R)$, respectively, and
$M_E$ is a mass matrix of hypothetical heavy charged leptons
$E_i$.

Stimulated by the successful derivation \cite{Koide06-S3} of 
the VEV relation (1.3), in the present paper, we will investigate
possible seesaw mass matrix structures  of the quarks and leptons 
$$
M_f = m_L M_F^{-1} m_R^\dagger ,
\eqno(1.6)
$$
by introducing heavy fermions 
$10'_{Li} +\overline{10}'_{Li} +1'_{Li}$  ($i=1,2,3$) of SU(5) 
in addition to the conventional quarks and leptons 
$\bar{5}_{Li} +10_{Li}$ as shown in Fig.~1.
Here, we assume that the VEVs of the flavor-triplet Higgs scalars 
$5_{Hi} +\bar{5}_{Hi} +1_{Hi}$ have the same structures which satisfy
the relations (1.3).
We consider that a variety of the mass spectra and mixings of quarks 
and neutrinos is caused by a variety of the structures of the heavy 
fermion mass matrices $M_F$.
As suggested by the feature (c), we want that the mass scale of $M_F$ 
is as low as possible.
We will build a seesaw mass matrix model with $M_F$ of the order of 10 TeV
for the quark sectors.

Note that although the relation (1.3) is a motivation for 
investigating the present model (Sec.~2), it is not essential in the
present paper whether the relation (1.3) is a fundamental law
or merely accidental.
The purpose of the present paper is to demonstrate that we can
explain the observed neutrino (and also quark) masses and mixings  
with the same values as the parameter values $v_i$ which are fixed by
$$
\frac{v_1}{\sqrt{m_e}}= \frac{v_2}{\sqrt{m_\mu}}=
\frac{v_3}{\sqrt{m_\tau}}=\frac{1}{\sqrt{m_e+m_\mu+m_\tau}} ,
\eqno(1.7)
$$
in the charged lepton sector.
In the present paper, we do not inquire the origin of the values of $v_i$.
In the present standpoint, the relation (1.1) is a phenomenological fact,
but it is not a theoretical result.  When we accept a seesaw mass matrix
form (1.6) which 
is motivated from the empirical relation (1.1), our interest is in how 
we can obtain the reasonable values of the quark and neutrino masses 
and mixings under the seesaw mass matrices (1.6) with a universal 
structure of $m_L$ (and $m_R$), but with sector-dependent structures of 
$M_F$.
In Secs.~3 and 4, we will assume a permutation symmetry S$_3$ 
for the structures of $M_F$.

\vspace{5mm}
{\large\bf 2 \ Fundamental fermions and scalars}

Suggested by the features (a), (b) and (c) 
discussed in Sec.~1, 
in this section, we discuss the seesaw mass
matrix form (1.6) concretely by introducing some additional
heavy fermions. 
For convenience, we use notations and conventions 
in an SU(5) GUT model for fermions and Higgs scalars, 
although we do not consider a gauge unification.
If we consider the unification of the gauge coupling constants,
it will be badly spoiled 
because there are many new particles in the present model.
Nevertheless, we consider that the SU(5) scheme is useful for
the description of the Yukawa interactions.
In the present model, we have the following fermions and
Higgs scalars:
$$ 
(\overline{5}_L + 10_L)_{(+)} 
+ (1'_L + \overline{10}'_L +  10'_L)_{(-)} 
+ (1_H +\overline{5}_{H} + 5_H)_{(-)} +(1'_H)_{(+)}  , 
\eqno(2.1)
$$
where the indices $(\pm)$ denote transformation
properties of a discrete symmetry Z$_2$. 
[Here, $\overline{10}'_L$ and $\bar{5}_H$ are not
Hermitian conjugates of $10'_L$ and $5_H$, respectively,
and $\overline{10}'_L$ and $10'_L$ ($\bar{5}_H$ and $5_H$)
are completely different particles each other.]
Therefore, we have the following Yukawa interactions:
$ 10'_L \bar{5}_L \bar{5}_H$, $10'_L 10_L 5_H$,
$ 1'_L \bar{5}_L {5}_H$, $10_L \overline{10}'_L 1_H$,
$10'_L \overline{10}'_L 1'_H$, and $1'_L 1'_L 1'_H$.
Here and hereafter, for convenience, we denote interaction 
terms as if those are superfields.
However, if we take those SUSY partners
into consideration at a low energy scale, the SU(3) color force
cannot become asymptotically free.
Therefore, we use those SUSY notations as an expedient.   
In other words,  we assume the absence of
the supersymmetric partners of the fields (2.1) at a low
energy scale.

Our essential assumption is as follows: 
the Higgs potentials for the 
scalars $5_H$, $\bar{5}_H$ and $1_H$ with the same Z$_2$ 
charges have the same structure.
This suggests that the scalars $(\bar{5}_H +5_H +1_H)$ 
will belong to a same multiplet in a higher flavor symmetry.
(However, in the present paper, we will not go into the investigation
of such a higher symmetry.)
As a result, the VEVs of the scalars $5_H$, $\bar{5}_H$ and $1_H$ 
take the same structures of the VEVs
$$
\langle 5_{Hi} \rangle = v_u z_i , \ \ 
\langle \bar{5}_{Hi} \rangle = v_d z_i , \ \ 
\langle 1_{Hi} \rangle  = v_s z_i,
\eqno(2.2)
$$
where $z_i$ are normalized as $z_1^2+z_2^2+z_3^2=1$ and
they satisfy the relation (1.3), i.e. 
$$
z_1^2 + z_2^2 + z_3^2 = \frac{2}{3}
 \left( z_1 + z_2 + z_3 \right) ^2,
\eqno(2.3)
$$
at the low energy scale $\mu=M_Z$.
Hereafter, for numerical estimates of the neutrino and quark 
mass matrices, we will use the values of
$z_i$:
$$
\frac{z_1}{\sqrt{m_e}}= \frac{z_2}{\sqrt{m_\mu}}=
\frac{z_3}{\sqrt{m_\tau}}=\frac{1}{\sqrt{m_e+m_\mu+m_\tau}} ,
\eqno(2.4)
$$
i.e. $z_1=0.016473$, $z_2=0.23678$
and $z_3=0.97140$.

On the other hand, we assume that the couplings of those Higgs scalars 
with fermions are structure-less:
$$
y_u \sum_i 10'_{Li} 10_{Li} 5_{Hi}
+ y_d \sum_i 10'_{Li} \bar{5}_{Li} \bar{5}_{Hi}
+ y_\nu \sum_i 1'_{Li} \bar{5}_{Li} {5}_{Hi}
+y_s \sum_i \overline{10}'_{Li} 10_{Li} 1_{Hi}.
\eqno(2.5)
$$

For convenience, hereafter, we denote the fermion mass terms
$\mu_{10}10'_L \overline{10}'_L $ as
$$
10'_{L} (\mu_{10}) \overline{10}'_{L}  =
 U_{Li} (\mu_{10}^Q)_{ij} \bar{U}_{Lj}  
+  D_{Li} (\mu_{10}^Q)_{ij} \bar{D}_{Lj}  
+ U_{Ri}^c (\mu_{10}^U)_{ij} \bar{U}_{Rj}^c 
+  E_{Ri}^c (\mu_{10}^E)_{ij} \bar{E}_{Rj}^c  ,
\eqno(2.6)
$$
where we have denoted the heavy fermions as 
$10'_L = [(U_L, D_L), U_R^c, E_R^c]$ and 
$\overline{10}'_L = [(\bar{U}_L, \bar{D}_L), \bar{U}_R^c, 
\bar{E}_R^c]$.
(Note that $\bar{U}_L$ is not the Hermitian conjugate of $U_L$,
and so on.)
Then, from the seesaw diagrams shown in Fig.~1, we obtain 
the following quark and lepton mass matrices: 
$$
(M_e)_{ij} = y_d y_s v_d v_s  z_i (\mu_{10}^E)^{-1}_{ij} z_j ,
\eqno(2.7)
$$
$$
(M_d)_{ij} = y_d y_s v_d v_s z_i (\mu_{10}^Q)^{-1}_{ij} z_j ,
\eqno(2.8)
$$
$$
(M_u)_{ij} = y_u y_s v_u v_s z_i \left[ (\mu_{10}^Q)^{-1}
+(\mu_{10}^U)^{-1} \right]_{ij} z_j ,
\eqno(2.9)
$$
$$
(M_\nu)_{ij} = y_\nu^2 v_u^2 z_i (y_S \langle 1'_H \rangle)^{-1}_{ij} z_j .
\eqno(2.10)
$$
Here, we have supposed 
$$
\langle 1_H \rangle \sim 10^2 \ {\rm GeV}, \ \  \mu_{10} \sim 10^4\ 
{\rm GeV}, \ \ \langle 1'_H \rangle \sim 10^{14}\ {\rm GeV}.
\eqno(2.11)
$$
Although the scalar $1'_H$ can couple not only to $1'_L 1'_L$, but also
to $\overline{10}'_L 10'_L$, the contributions $\langle 1'_H \rangle^{-1}$
in the mass matrices $M_f$ ($f=e, u, d$) are negligibly small
compared with $(\mu_{10})^{-1}$.

In order to explain the relation (1.1), we must take
 $(\mu_{10}^E)_{ij} =\mu_{10} \delta_{ij}$ so that 
$$
(M_e)_{ij} = \frac{y_d v_d y_s v_s}{\mu_{10}} z_i \delta_{ij} z_j .
\eqno(2.12)
$$
At present, we do not inquire why the mass matrix $\mu_{10}^E$
is structure-less.
We consider that the matrices $M_F$ are, in general, not structure-less,
and their structures are dependent on the sectors, so that
mass spectra and mixings individual sectors appear. 

The present model is based on a multi-Higgs model, because
our Higgs scalars $5_H$ and $\bar{5}_H$ are flavor-triplets.
In general, such a model leads to a serious trouble, i.e. 
the flavor-changing neutral current (FCNC) problem.
However, since our Higgs scalars $5_H$ and $\bar{5}_H$ couple to
the quarks and leptons not directly, but via 
$10'_L \bar{5}_L \bar{5}_H$ and $10'_L 10_L 5_H$,
the FCNC problem in the present model can substantially be
evaded.
Roughly speaking, when we denote the mass matrices $M_u$ and
$M_d$ given in Eqs.~(2.8) and (2.9) as $M_q = m_L M_Q^{-1}m_s$
symbolically,
the effective interactions of $\overline{q}q$ with the Higgs 
scalars $\phi$ ($5_H$ or $\overline{5}_H$) are given by
$\overline{q} \phi M_Q^{-1} m_s q$, so that the effective FCNC
interactions through $\phi$ are suppressed by the order of
$(M_Q^{-1} m_s)^2 \sim 10^{-4}$.
Therefore, the FCNC effects practically become invisible.

\vspace{5mm}
{\large\bf 3 \ Quark mass matrices}

In order to obtain realistic quark mass matrices, 
we must consider that the quark sectors in the heavy 
fermion mass terms (2.6)  have some structures
differently from the lepton sector $\mu_{10}^E$.
The Yukawa interaction (2.6) is invariant under 
a permutation symmetry S$_3$.
(The form (2.6) is not a general form of the S$_3$
invariant Yukawa interactions. The form is  
constrained more than the S$_3$ symmetry.)
Therefore, we assume that the heavy fermion mass matrices
 $\mu_{10}^F$  ($F=Q,U$) are also S$_3$-invariant.
We concretely assume that  $\mu_{10}^F$ ($F=Q,U$) 
are diagonal on the $(F_\pi, F_\eta, F_\sigma)$ basis, i.e.
$$
M_D \left( \overline{F}_\pi F_\pi +\overline{F}_\eta F_\eta \right)
+M_S \overline{F}_\sigma F_\sigma  ,
\eqno(3.1)
$$
after the SU(5) symmetry was broken,
where $F_\sigma$ and $(F_\pi, F_\eta)$ are singlet and doublet of
S$_3$, respectively, and those are define by
$$
\left(
\begin{array}{c}
F_\pi \\
F_\eta \\
F_\sigma 
\end{array} \right)
=
A \left(
\begin{array}{c}
F_1 \\
F_2 \\
F_3 
\end{array} \right)
\eqno(3.2)
$$
$$
A=\left( \begin{array}{ccc}
\frac{1}{\sqrt{2}} & -\frac{1}{\sqrt{2}} & 0 \\
\frac{1}{\sqrt{6}} & \frac{1}{\sqrt{2}} & -\frac{2}{\sqrt{6}} \\
\frac{1}{\sqrt{3}} & \frac{1}{\sqrt{3}} & \frac{1}{\sqrt{3}}
\end{array} \right) .
\eqno(3.3)
$$
Then, the inverse of the heavy fermion mass matrices, $(\mu_{10}^F)^{-1}$, 
are given by
$$
(\mu_{10}^F)^{-1} = A^T {\rm diag}(M_D^{-1}, M_D^{-1}, M_S^{-1}) A
$$
$$
= \frac{1}{M_D} ({\bf 1}-X) +\frac{1}{M_S} X ,
\eqno(3.4)
$$
where
$$
{\bf 1}=\left(
\begin{array}{ccc}
1 & 0 & 0 \\
0 & 1 & 0 \\
0 & 0 & 1
\end{array} \right) , \ \ \ 
X= \frac{1}{3} \left(
\begin{array}{ccc}
1 & 1 & 1 \\
1 & 1 & 1 \\
1 & 1 & 1
\end{array} \right) . 
\eqno(3.5)
$$
In other words, the fermion mass matrices $\mu_{10}^F$  
are given by the following S$_3$-invariant form
$$
(\mu_{10}^F)_{ij} = \mu_{10} \left( {\bf 1}_{ij} + 3 b_F X_{ij}
\right),
\eqno(3.6)
$$
where $3 b_F =M_D^F/M_S^F-1$.
The case in the charged lepton sector corresponds to a specific case
with $M_D=M_S$, i.e. a case of $b_F=0$.

The quark mass matrices $M_u$ and $M_d$ with the forms (3.6)
have already investigated in Ref.~\cite{DUSM}
as the so-called ``democratic universal seesaw mass matrix model",
where it has been found that the values $b_u= -1/3$ and 
$b_d = -e^{i\beta}$ ($\beta\simeq 20^\circ$) can give reasonable 
quark masses and the Cabibbo-Kobayashi-Maskawa (CKM) mixing matrix 
parameters.
In the present model, the parameters $(b_d, b_u)$ in Ref.~\cite{DUSM}
correspond to $(b_Q, b_U)$, so that we take 
$$
b_Q \simeq - e^{i\beta} \ \ (\beta \simeq 20^\circ), \ \ \ 
b_U \simeq -\frac{1}{3} .
\eqno(3.7)
$$
As pointed out in Ref.~\cite{DUSM}, for the choice $b_U=-1/3$,
there is no inverse matrix of $\mu_{10}^U$ (i.e. det$(\mu_{10}^U)=0$),
so that one of the up-quark masses has a mass of the
order of $v_u$, and we identify it as the top quark mass.
Another prediction from $b_U=-1/3$ is \cite{Koide93,DUSM}
$$
\frac{m_u}{m_c} \simeq \frac{3 m_e}{4 m_\mu}.
\eqno(3.8)
$$
Also, the choice $b_Q \simeq -1$ leads to the relations \cite{DUSM}
$$
\frac{m_c}{m_b} \simeq 4 \frac{m_\mu}{m_\tau}, \ \ 
\frac{m_d m_s}{m_b^2} \simeq 4 \frac{m_e m_\mu}{m_\tau^2}, \ \ 
\frac{m_u}{m_d} \simeq 3 \frac{m_s}{m_c} \simeq 3 
\left|\sin\frac{\beta}{2}\right|. 
\eqno(3.9)
$$ 
Since the purpose of the present model is to give the outline of
the model, we do not give numerical re-fitting of the values $(b_Q, b_U)$.
We also do not reject a possibility that there is another parameter set
$(b_Q, b_U)$ which leads to favorable quark masses and CKM parameters.

\vspace{5mm}
{\large\bf 4 \ Neutrino mass matrix}

In the present section, we investigate
whether the model given in Sec.~2 
in order to understand the charged lepton mass formula
(1.1) can explain the observed neutrino masses and
mixings or not.
Obviously, if $\langle 1'_H\rangle$ in Eq.~(2.10) is also
structure-less, i.e. if $\langle 1'_H\rangle$ has a unit
matrix form, the neutrino mass matrix $M_\nu$ becomes
a diagonal mass matrix as well as the charged lepton
mass matrix $M_e$, so that we can obtain neither neutrino
mixings nor reasonable neutrino mass spectrum.
Also, if we assume that $M_R = y_S \langle 1'_H\rangle$
has the same structure as $M_F$ (3.6) in the quark
sectors, we will find that such a model cannot explain
the observed neutrino data \cite{Koide-nu}.
The purpose of the present section is to investigate
what additional assumptions are needed for the 
explanation of the observed neutrino data.

In the expression (2.10) of the neutrino mass matrix $M_\nu$,
we have already assumed that the heavy fermion mass terms 
$\mu_1 1'_L 1'_L$ are sufficiently large to be neglected 
compared with the contribution of $\langle 1'_H \rangle$.
We consider the observed peculiar structure of the neutrino mass
matrix comes from the interactions among the heavy particles, 
$ 1'_{L} 1'_{L} 1'_{H}$.
We  assume 
the following simple S$_3$-invariant form for the 
interactions $1'_{L} 1'_{L} 1'_{H}$:
$$
(y_S)_{ijk} 1'_{Li} 1'_{Lj} 1'_{Hk}  = y_S
(1'_{L1}\ 1'_{L2}\ 1'_{L3}) \left(
\begin{array}{ccc}
1'_{H1} & 1'_{H3} & 1'_{H2} \\
1'_{H3} & 1'_{H2} & 1'_{H1} \\
1'_{H2} & 1'_{H1} & 1'_{H3}
\end{array} \right)
\left(
\begin{array}{c}
1'_{L1} \\
 1'_{L2} \\
 1'_{L3}
\end{array} \right) .
\eqno(4.1)
$$
(Of course, the form (4.1) is not a general form of
the S$_3$-invariant cubic interactions.
Only when we require both a cyclic permutation
symmetry and the S$_3$ symmetry, the possible
forms of the Yukawa interactions are confined
in the two forms (2.5) and (4.1).)
When we denote the VEVs of $1'_H$ as 
$\langle 1'_{Hi} \rangle = v_S Z_i$ 
with a normalization condition $Z_1^2+Z_2^2+Z_3^2=1$,
from Eq.~(2.10), we obtain the neutrino mass matrix 
$$
M_\nu = m_0 \left(
\begin{array}{ccc}
z_1^2 (Z_1^2 -Z_2 Z_3) & z_1 z_2 (Z_3^2 -Z_1 Z_2)
& z_1 z_3 (Z_2^2 -Z_1 Z_3) \\
z_1 z_2 (Z_3^2 -Z_1 Z_2) & z_2^2 (Z_2^2 -Z_1 Z_3)
& z_2 z_3 (Z_1^2 -Z_2 Z_3) \\
z_1 z_3 (Z_2^2 -Z_1 Z_3) & z_2 z_3 (Z_1^2 -Z_2 Z_3) 
& z_3^2 (Z_3^2 -Z_1 Z_2)
\end{array} \right) ,
\eqno(4.2)
$$
where $m_0 = (y_\nu^2 v_u^2/y_S v_S)/(Z_3^3+Z_2^3+Z_1^3-3 Z_1 Z_2 Z_3)$.
In order to obtain a nearly bimaximal mixing, we must be take
$z_2 Z_2 \simeq z_3 Z_3$.

The parameters $Z_i$ are free from the values $z_i$,
because the VEVs $\langle 1'_{H} \rangle$ may have 
different values from the VEVs of $(\bar{5}_H +5_H +1_H)$.
However, from  an economical point of view of the parameters,
we interest in a case that the parameters $Z_i$ also
satisfy the relation (1.3) as well as $z_i$.
Considering the phenomenological requirement 
$z_2 Z_2 \simeq z_3 Z_3$,  by way of trial, we assume 
$$
(Z_1, Z_2, Z_3) = (z_1, z_3, z_2).
\eqno(4.3)
$$
Since the energy scales of $\langle 1'_H \rangle$ and
$\langle 1_H \rangle$ are different from each other
(i.e. $\langle 1'_H \rangle \sim 10^{14}$ GeV and
$\langle 1_H \rangle \sim 10^2$ GeV), we do not consider
that the relation (4.3) is exact at a low energy scale.  

At present, we do not know the origin of such the inversion 
$2\leftrightarrow 3$,
and it is a pure phenomenological assumption.
Although we can obtain favorable predictions of the neutrino
masses and mixings for the trial choice (4.3)  as we show below, 
we can also obtain
favorable results for suitable parameter values of
$(Z_1,Z_2,Z_3)$ without the assumption (4.3).
The choice (4.3) is merely one of the successful parameter
values $Z_i$.
The relation (4.3) is not an essential assumption in the
present model.

For the trial choice (4.3), we find the following numerical
results:
$$
m_{\nu 1}=0.00737 m_0, \ \  m_{\nu 2}=0.01651 m_0, \ \  
m_{\nu 3}=0.09965 m_0, 
\eqno(4.4)
$$
$$
U = \left(
\begin{array}{ccc}
0.8011 & -0.5904 & 0.0985 \\
0.4532 & 0.4907  & -0.7442 \\
0.3911 & 0.6408 & 0.6607 
\end{array}  \right) ,
\eqno(4.5)
$$
so that we obtain
$$
R = \frac{\Delta m^2_{21}}{\Delta m^2_{32}} = 0.023 ,
\eqno(4.6)
$$
$$
\sin^2 2\theta_{23} = 0.97 ,
\eqno(4.7)
$$
$$
\tan^2 \theta_{12} = 0.54 ,
\eqno(4.8)
$$ 
which are in good agreement with the present observed best-fit 
values \cite{Kamland,K2K} 
$R \simeq (7.9 \times 10^{-5})/(2.8\times 10^{-3})=0.029$, 
$\sin^2 2\theta_{atm}=1.0$ and 
$\tan^2 \theta_{solar}=0.40^{+0.10}_{-0.07}$.
It is worth noticing that we do not have any free
parameter in the neutrino sector except for the postulation
(4.3).
Of course, if we take slight deviations from the 
assumption (4.3), we can obtain more
excellent agreements with the observed values.
Thus, at least, we can say that we are going in the right direction.

If we put $m_{\nu 3} = \sqrt{\Delta m^2_{atm}}=0.053$ eV, then we obtain
$$
m_{\nu 1}=0.0039\ {\rm eV}, \ \  m_{\nu 2}=0.0088\ {\rm eV}, \ \  
m_{\nu 3}=0.053\ {\rm eV}. 
\eqno(4.9)
$$

\vspace{5mm}
{\large\bf 5 \ Concluding remarks}

The purpose of the present paper is not to explain the
charged lepton mass formula (1.1).  When we consider that the
relation is remarkably satisfied at a low energy scale,
we inevitably reach to the idea that the mass spectrum 
originates not in the Yukawa coupling structure at a 
unification energy scale, but in the VEV structure of a 
three-flavor Higgs structure at the low energy scale.  
The purpose of the present paper is also not to investigate 
the validity of the mass matrix (1.3) for the charged leptons.
Our interest is in an extension of the mass matrix (1.3) 
to mass matrices of the quarks and neutrinos. 
The purpose of the present paper is to investigate whether 
such a model can also explain or not the observed quark and 
neutrino masses and mixings with the universal structure of
$m_L$ ($m_R$) which is fixed in the charged lepton sector. 

In Sec.~2, we have assumed the additional fermions
$10'_L +\overline{10}'_L +1'_L$. 
If we consider another models, for example, with 
$5'_L+\overline{5}'_L$, 
we will encounter some troubles when we try to build
a universal seesaw model for quarks and neutrinos.
Only the choice $10'_L +\overline{10}'_L +1'_L$
yields a natural extension of the seesaw mass matrix
model for charged leptons (with $M_E \sim 10^4$ GeV) 
to a model for the quarks and leptons.
The essential assumption in Sec.~2 is Eq.~(2.2), i.e.
the VEV structures of the scalars $5_H +\overline{5}_H +1_H$
are universal.
Then, our interest was whether such a model can also explain
the observed quark and neutrino masses and mixings or not.

For quark sectors, we have assumed that the heavy fermion
mass terms are invariant under the S$_3$ symmetry, i.e. 
the heavy fermion mass matrices $M_F=\mu_{10}^F$ take 
the form (3.1), which leads to  the  democratic seesaw mass matrix 
form (3.4).
We can find the parameter values which can give reasonable
quark masses and CKM mixing parameters.

For the neutrino sector, our essential assumption is 
the S$_3$-invariant interactions (4.1) of the heavy fermions
$1'_L$.
Then, the parameters $Z_i=\langle 1'_{Hi}\rangle$ can take
values which can give reasonable values of the neutrino 
masses and mixings.
Especially, it is interesting that the values also satisfy
the relation (1.3) as well as $v_i=\langle \phi_{i}\rangle$ 
($\phi_i = 5_{Hi}$, $\overline{5}_{Hi}$ and $1_{Hi}$).
The choice $(Z_1, Z_2, Z_3)=(z_1, z_3, z_2)$, Eq.~(4.3), 
can give $\Delta m^2_{solar}/\Delta m^2_{atm}=0.023$,
$\sin^2 2\theta_{atm}=0.97$, and $\tan^2\theta_{solar}=0.54$.
(Of course, those are not inevitable predictions in the
present model. The choice (4.3) is merely an example of the
parameter choice.)

The present model, at present, can give neither gauge unification 
nor SUSY scenario.
However, we may say that the three-flavor Higgs model 
with the VEVs $v_i \propto \sqrt{m_{ei}}$ has a possibility
to explain quark and lepton mass matrices with the same
parameter values of $v_i$.
The investigation of a possibility that the fermion masses
are closely related to VEVs of three-flavor Higgs scalars
is just in the beginning.

\vspace{5mm}
\centerline{\bf Acknowledgments}

The author is also grateful to Professors J.~Kubo and 
D.~Suematsu, and to young Particle Physics members at 
Kanazawa University for enjoyable and helpful
discussions. 
He also thanks Dr.~S.~Kaneko for pointing out a numerical
error in an earlier version of this work.


\newpage
\vspace{-2cm}
\hspace*{4cm}
\begin{picture}(300,120)(50,50)
\put(0,50){\thicklines \vector(1,0){40}}
\put(40,50){\thicklines \line(1,0){35}}
\put(35,30){$\bar{5}_{L}$}
\put(75,50){\thicklines \line(0,1){5}}
\put(75,60){\thicklines \line(0,1){5}}
\put(75,70){\thicklines \line(0,1){5}}
\put(75,80){\thicklines \line(0,1){5}}
\put(70,85){\thicklines \line(1,1){10}}
\put(80,85){\thicklines \line(-1,1){10}}
\put(75,50){\circle*{5}}
\put(65,105){$\langle \bar{5}_H \rangle$}
\put(150,50){\thicklines \vector(-1,0){40}}
\put(110,50){\thicklines \line(-1,0){35}}
\put(115,30){$10'_{L}$}
%
\put(145,70){$\mu_{10}$}
\put(145,45){\thicklines \line(1,1){10}}
\put(155,45){\thicklines \line(-1,1){10}}
\put(150,50){\thicklines \vector(1,0){40}}
\put(190,50){\thicklines \line(1,0){35}}
\put(185,30){$ \overline{10}'_{L} $}
\put(225,50){\thicklines \line(0,1){5}}
\put(225,60){\thicklines \line(0,1){5}}
\put(225,70){\thicklines \line(0,1){5}}
\put(225,80){\thicklines \line(0,1){5}}
\put(220,85){\thicklines \line(1,1){10}}
\put(230,85){\thicklines \line(-1,1){10}}
\put(225,50){\circle*{5}}
\put(215,105){$\langle 1_H \rangle$}
\put(300,50){\thicklines \vector(-1,0){40}}
\put(260,50){\thicklines \line(-1,0){35}}
\put(265,30){$10_{L}$}
\end{picture}

\vspace*{1cm}
\begin{center}
{\small\bf  (a) $M_e$ and $M_d$ }
\end{center}

\vspace{-2cm}
\hspace*{4cm}
\begin{picture}(300,120)(50,50)
\put(0,50){\thicklines \vector(1,0){40}}
\put(40,50){\thicklines \line(1,0){35}}
\put(35,30){$10_{L}$}
\put(75,50){\thicklines \line(0,1){5}}
\put(75,60){\thicklines \line(0,1){5}}
\put(75,70){\thicklines \line(0,1){5}}
\put(75,80){\thicklines \line(0,1){5}}
\put(70,85){\thicklines \line(1,1){10}}
\put(80,85){\thicklines \line(-1,1){10}}
\put(75,50){\circle*{5}}
\put(65,105){$\langle 5_H \rangle$}
\put(150,50){\thicklines \vector(-1,0){40}}
\put(110,50){\thicklines \line(-1,0){35}}
\put(115,30){$10'_{L}$}
%
\put(145,65){$\mu_{10}$}
\put(145,45){\thicklines \line(1,1){10}}
\put(155,45){\thicklines \line(-1,1){10}}
\put(150,50){\thicklines \vector(1,0){40}}
\put(190,50){\thicklines \line(1,0){35}}
\put(185,30){$ \overline{10}'_{L} $}
\put(225,50){\thicklines \line(0,1){5}}
\put(225,60){\thicklines \line(0,1){5}}
\put(225,70){\thicklines \line(0,1){5}}
\put(225,80){\thicklines \line(0,1){5}}
\put(220,85){\thicklines \line(1,1){10}}
\put(230,85){\thicklines \line(-1,1){10}}
\put(225,50){\circle*{5}}
\put(215,105){$\langle 1_H \rangle$}
\put(300,50){\thicklines \vector(-1,0){40}}
\put(260,50){\thicklines \line(-1,0){35}}
\put(265,30){$10_{L}$}
\end{picture}

\vspace*{1cm}
\begin{center}
{\small\bf (b) $M_u$
 }
\end{center}

\vspace{-2cm}
\hspace*{4cm}
\begin{picture}(300,120)(50,50)
\put(0,50){\thicklines \vector(1,0){40}}
\put(40,50){\thicklines \line(1,0){35}}
\put(35,30){$\bar{5}_{L}$}
\put(75,50){\thicklines \line(0,1){5}}
\put(75,60){\thicklines \line(0,1){5}}
\put(75,70){\thicklines \line(0,1){5}}
\put(75,80){\thicklines \line(0,1){5}}
\put(70,85){\thicklines \line(1,1){10}}
\put(80,85){\thicklines \line(-1,1){10}}
\put(75,50){\circle*{5}}
\put(65,105){$\langle 5_H \rangle$}
\put(150,50){\thicklines \vector(-1,0){40}}
\put(110,50){\thicklines \line(-1,0){35}}
\put(115,30){$1'_{L}$}
\put(150,50){\circle*{5}}
\put(150,50){\thicklines \line(0,1){5}}
\put(150,60){\thicklines \line(0,1){5}}
\put(150,70){\thicklines \line(0,1){5}}
\put(150,80){\thicklines \line(0,1){5}}
\put(145,85){\thicklines \line(1,1){10}}
\put(155,85){\thicklines \line(-1,1){10}}
\put(150,50){\circle*{5}}
\put(140,105){$\langle 1'_H \rangle$}
\put(150,50){\thicklines \vector(1,0){40}}
\put(190,50){\thicklines \line(1,0){35}}
\put(185,30){$ 1'_{L} $}
\put(225,50){\thicklines \line(0,1){5}}
\put(225,60){\thicklines \line(0,1){5}}
\put(225,70){\thicklines \line(0,1){5}}
\put(225,80){\thicklines \line(0,1){5}}
\put(220,85){\thicklines \line(1,1){10}}
\put(230,85){\thicklines \line(-1,1){10}}
\put(225,50){\circle*{5}}
\put(215,105){$\langle 5_H \rangle$}
\put(300,50){\thicklines \vector(-1,0){40}}
\put(260,50){\thicklines \line(-1,0){35}}
\put(265,30){$\bar{5}_{L}$}
\end{picture}

\vspace*{1cm}
\begin{center}
{\small\bf (c) $M_\nu$ 
 }
\end{center}

\begin{quotation}
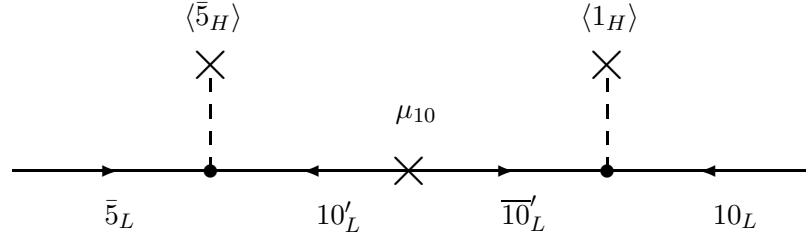
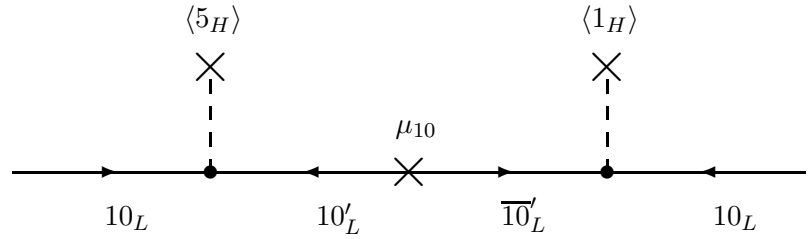
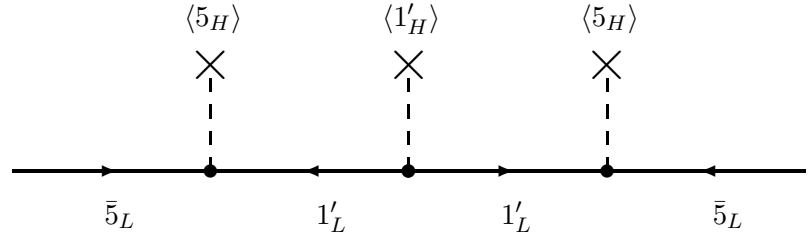
{\bf Fig.~1 Seesaw mass generation of the quark and leptons: 
(a) charged lepton and down-quark mass matrices $M_e$ and $M_d$,
(b) up-quark mass matrix $M_u$ and (c) neutrino mass matrix
$M_\nu$.}
\end{quotation}

\vspace{3mm}


\begin{thebibliography}{99}
%
%
%
\bibitem{Koide82} Y.~Koide, Lett.~Nuovo Cimento {\bf 34}, 201 (1982).
%
\bibitem{Koide83} Y.~Koide, Phys.~Rev. {\bf D28}, 252 (1983).
%
\bibitem{Koide90} Y.~Koide, Mod.~Phys.~Lett. {\bf A5}, 2319 (1990).
%
%
%
\bibitem{PDG04}
S.~Eidelman {\it et al.} (Particle Data Group), 
Phys.~Lett. {\bf B592}, 1 (2004). 
%
\bibitem{UnivSeesaw} The seesaw model for charged particles is known
as the ``universal seesaw model":
Z.~G.~Berezhiani, Phys.~Lett.~{\bf 129B}, 99 (1983);
Phys.~Lett.~{\bf 150B}, 177 (1985);
D.~Chang and R.~N.~Mohapatra, Phys.~Rev.~Lett.~{\bf 58},1600 (1987); 
A.~Davidson and K.~C.~Wali, Phys.~Rev.~Lett.~{\bf 59}, 393 (1987);
S.~Rajpoot, Mod.~Phys.~Lett. {\bf A2}, 307 (1987); 
Phys.~Lett.~{\bf 191B}, 122 (1987); Phys.~Rev.~{\bf D36}, 1479 (1987);
K.~B.~Babu and R.~N.~Mohapatra, Phys.~Rev.~Lett.~{\bf 62}, 1079 (1989); 
Phys.~Rev. {\bf D41}, 1286 (1990); 
S.~Ranfone, Phys.~Rev.~{\bf D42}, 3819 (1990); 
A.~Davidson, S.~Ranfone and K.~C.~Wali, 
Phys.~Rev.~{\bf D41}, 208 (1990); 
I.~Sogami and T.~Shinohara, Prog.~Theor.~Phys.~{\bf 66}, 1031 (1991);
Phys.~Rev. {\bf D47}, 2905 (1993); 
Z.~G.~Berezhiani and R.~Rattazzi, Phys.~Lett.~{\bf B279}, 124 (1992);
P.~Cho, Phys.~Rev. {\bf D48}, 5331 (1994); 
A.~Davidson, L.~Michel, M.~L,~Sage and  K.~C.~Wali, 
Phys.~Rev.~{\bf D49}, 1378 (1994); 
W.~A.~Ponce, A.~Zepeda and R.~G.~Lozano, 
Phys.~Rev.~{\bf D49}, 4954 (1994).
%
%
%
%
%
%
\bibitem{KF96} Y.~Koide and H.~Fusaoka, Z.~Phys. {\bf C71},
459 (1996).
%
\bibitem{KT96} Y.~Koide and M.~Tanimoto,  Z.~Phys. {\bf C72},
333 (1996).
%
\bibitem{Koide99} Y.~Koide, Phys.~Rev. {\bf D60}, 077301 (1999).
%
%
%
\bibitem{S3} S.~Pakvasa and H.~Sugawara, Phys.~Lett. {\bf B73}, 61
(1978); H.~Harari, H.~Haut and J.~Weyers, Phys.~Lett. {\bf B78},
459 (1978); E.~Derman, Phys.~Rev. {\bf D19}, 317 (1979);
D.~Wyler, Phys.~Rev. {\bf D19}, 330 (1979).
%
%
\bibitem{Koide06-S3} Y.~Koide, hep-ph/0509214, Phys.~Rev. {\bf D73}, 057901 
(2006).
%
\bibitem{DUSM} 
Y.~Koide and H.~Fusaoka, Z.~Phys. {\bf C71}, 459 (1996); 
Prog.~Theor.~Phys. {\bf 97}, 459 (1997).
%
%
\bibitem{Koide93} Y.~Koide, Mod.~Phys.~Lett. {\bf A8}, 2071 (1993).
%
%
\bibitem{Koide-nu} Y.~Koide, Mod.~Phys.~Lett. {\bf A11}, 2849 (1996);
Phys.~Rev. {\bf D57},  5836 (1998).
%
%
\bibitem{Kamland} T.~Araki, {\it et al}., KamLAND Collaboration,
Phys.~Rev.~Lett. {\bf 94}, 081801 (2005).
%
\bibitem{K2K} E.~Aliu,  {\it et al}., K2K Collaboration,
Phys.~Rev.~Lett. {\bf 94}, 081802 (2005).
\end{thebibliography}
\end{document}